\documentstyle[prl,aps,twocolumn,floats,amssymb,epsfig]{revtex}

\def\bbR{{\mathbb R}}

\def\su{{\rm{su}}}
\def\u{{\rm{u}}}

\def\U{{\rm{U}}}

\def\SO{{\rm{SO}}}
\def\O{{\rm{O}}}

\newcommand{\mb}[1]{\ifmmode#1\else\mbox{$#1$}\fi}

\newcommand\al{\mb{\alpha}}

\newcommand\be{\mb{\beta}}

\newcommand\la{\mb{\lambda}}

\newcommand\calG{\mb{{\cal G}}}
\newcommand\calH{\mb{{\cal H}}}

\newcommand\calL{\mb{{\cal L}}}
\newcommand\calM{\mb{{\cal M}}}
\newcommand\calN{\mb{{\cal N}}}
\newcommand\calO{\mb{{\cal O}}}

\newcommand\calT{\mb{{\cal T}}}

\newcommand\calV{\mb{{\cal V}}}

\newcommand{\beq}{\begin{equation}}
\newcommand{\eeq}{\end{equation}}
\newcommand{\nn}{\nonumber}
\newcommand{\bea}{\begin{eqnarray}}
\newcommand{\eea}{\end{eqnarray}}
\newcommand{\norm}[1]{\parallel \! {#1} \! \parallel}

\newcommand{\inprod}[2]{\langle {#1}, {#2} \rangle}

\newcommand{\emb}{\mb{{\rm emb}}}
\newcommand{\tr}{\mb{{\rm tr}\,}}

\newcommand{\pderiv}[2]{\frac{\partial {#1}}{\partial {#2}}}
\newcommand{\x}{\mb{\times}}
\newcommand{\rhat}{\mb{\hat{\bm r}}}

\newcommand{\Ad}{{\rm Ad}}
\newcommand{\ad}{{\rm ad}}
\newcommand{\rank}{{\rm rank}}
\newcommand{\gsim}
{\raise.3ex\hbox{$\;>$\kern-.75em\lower1ex\hbox{$\sim$}$\:$}}
\newcommand{\lsim}
{\raise.3ex\hbox{$\;<$\kern-.75em\lower1ex\hbox{$\sim$}$\:$}}
\newcommand{\ts}{\textstyle}
\newcommand{\ds}{\displaystyle}
\newcommand{\half}{\mb{\ts \frac{1}{2}}}
\newcommand{\quarter}{\mb{\ts \frac{1}{4}}}

\newcommand{\bm}[1]{{\mbox{\boldmath $#1$}}}

\begin{document}
\renewcommand{\thesubsection}{\arabic{subsection}} 
\draft


\twocolumn[\hsize\textwidth\columnwidth\hsize\csname @twocolumnfalse\endcsname
\title{Embedded monopoles}
\author{Nathan\  F.\  Lepora\footnotemark}
\date{Submitted 26 October 2002; accepted 14 March 2002}
\maketitle

\begin{abstract}
Using the embedded defect method, we classify the
possible embeddings of a 't~Hooft-Polyakov monopole in a general
gauge theory. We then discuss some similarities with embedded vortices 
and relate our results to fundamental monopoles.
\end{abstract}
\pacs{Published in: Phys. Lett. {\bf B}533 (2002) 137.}
]


\subsection{Introduction}

When describing classical solutions to spontaneously broken
Yang-Mills theories, there are several available methods. If the solutions
are topologically stable, there is a topological 
classification~\cite{tom}; while 
underneath this structure, there are finer classifications relating to 
the degeneracy of classical solutions. Such degeneracy is important for
properties other than stability --- for instance,
Goddard, Nuyts and Olive use such a classification of non-Abelian
monopoles to conjecture a dual gauge group~\cite{gno}.

For non-Abelian monopoles, a finer than topological classification can be
achieved by using fundamental monopoles~\cite{wein}. Any 
monopole can be described as a composite of several fundamental monopoles, 
each of which is associated with a simple root of the gauge group.
Furthermore, in the BPS limit each fundamental monopole 
is known to take an Prasad-Sommerfeld form on the su(2) algebra 
defined by its root~\cite{wein}.

For defects in general (BPS and non-BPS),
a good formalism to describe a finer than topological
classification is the embedded defect method of Barriola, Vachaspati,
and Bucher~\cite{vach94}. This method describes how a solution in one 
gauge theory is embedded in a larger theory and gives some
constraints for when this embedding is allowed. Of particular interest are
embedded vortices and embedded monopoles.

For embedded vortices, the embedding conditions of 
ref.~\cite{vach94} can be expressed in a very simple way that directly relates
to the geometry of the vacuum manifold~\cite{mevor}. 
It can then be shown that embedded vortices separate into classes under
the action of the residual gauge symmetry and from their embeddings 
easily specify their gauge degeneracy.

Such a picture has not been applied to embedded monopoles.
For that reason, this letter aims to:\\
(a) Apply embedded defect methods to monopoles.\\
(b) Find similarities between embedded monopoles and
embedded vortices.\\
(c) Describe how these results on embedded monopoles
relate to fundamental monopoles.

In addition, this gives some new results and a new picture of
non-Abelian monopoles, both of which may be useful when examining
their properties.

\subsection{Formalism}
Consider a spontaneously broken Yang-Mills theory with a
compact semi-simple gauge group $G$ and a scalar field $\Phi\in \calV$
in the $D$ representation of $G$
\beq
\label{lag}
\calL[\Phi,A^\mu] = -\quarter \inprod{F_{\mu\nu}}{F^{\mu\nu}} + 
\half \inprod{D_\mu\Phi}{D^\mu\Phi} - V[\Phi].
\eeq
Here the covariant derivative is
$D^\mu\Phi = \partial^\mu\Phi + d(A^\mu)\Phi$, while the
field tensor is
$F^{\mu\nu} = \partial^\mu A^\nu - \partial^\nu A^\mu + [A^\mu,A^\nu]$,
a covariant curl. The derived representation $d$
is defined by $e^{d(X)}=D(e^X)$, where $X$ is in the Lie algebra $\calG$ 

In this letter, we use a coordinate independent notation, which
we think better reflects the geometry behind most of our
results~\cite{mevor,megeom}. 
Then the gauge kinetic term in (\ref{lag}) is
defined by an inner product on $\calG$
\beq
\label{XY}
\inprod{X}{Y} = -\frac{2}{f_a^2}\, \tr [\ad(X)\ad(Y)],\hspace{2em}
X,Y\in\calG_a,
\eeq
with $f_a$ a coupling constant for each simple subgroup 
$G_a\subseteq G$.
Likewise, the scalar kinetic term is defined by a Euclidean inner
product $\inprod{\Phi_1}{\Phi_2}$ on $\calV$~\cite{com}.

In this notation, the field equations are
\bea
\label{fe}
D_\mu D^\mu\Phi=-\pderiv{V}{\Phi},\hspace{1.5em}
D_\mu F^{\mu\nu}=J^\nu,
\eea
with current 
$\inprod{J^\nu}{Y}=\inprod{d(Y)\Phi}{D^\nu\Phi}-\inprod{D^\nu\Phi}{d(Y)\Phi}$
and covariant derivative
$D_\mu F^{\mu\nu}= \partial_\mu F^{\mu\nu}+[A_\mu, F^{\mu\nu}]$.

If the potential $V[\Phi]$ is minimized at some value $\Phi_0$,
the residual gauge symmetry is
\beq
\label{H=}
H = \left\{h\in G:D(h)\Phi_0=\Phi_0\right\},
\eeq
with $\calH$ the Lie algebra of $H$.
Then (\ref{H=}) defines an $\Ad(H)$-invariant decomposition of $\calG$
into massless and massive gauge boson generators
\beq
\label{reddec}
\calG = \calH \oplus \calM,\hspace{2em}
\Ad(H)\calH \subseteq \calH,\hspace{.5em}
\Ad(H)\calM \subseteq \calM.
\eeq
Here $\Ad$ refers to the adjoint representation of $G$ on $\calG$,
where $\Ad(g)X=gXg^{-1}$ and the derived representation is
$\ad(X)Y=[X,Y]$ for $X,Y\in\calG$.

A central feature of this and some related papers~\cite{mevor,megeom} is the 
reduction of $\calM$ into irreducible subspaces under the adjoint action 
of $H$. These subspaces correspond to irreducible representations of $H$ 
on $\calM$:
\beq
\label{Mdecom}
\calM= \calM_1 \oplus \cdots \oplus \calM_n.
\eeq
Physically, each $\calM_a$ defines a gauge multiplet of 
massive gauge bosons (for instance, the W and Z-bosons in electroweak theory).

Embedded defects~\cite{vach94}
are $(2-k)$-dimensional topological defects that remain a solution
when they are embedded into a larger gauge theory
(with $k=0,1,2$ for domain walls, vortices and monopoles). 
They are defined by an inclusion of their gauge symmetry breaking in
that of the full theory
\bea
\label{emb}
G\hspace{1em} &\rightarrow& \ H \nn\\
\cup\hspace{1em} && \hspace{.5em}\cup\hspace{4em}
\pi_k (G_\emb / H_\emb)\neq 0.\\
G_\emb &\rightarrow& \ H_\emb \nn
\eea

In the original reference~\cite{vach94}, these embedded defects have fields
fully embedded over the spatial domain
\beq
\label{embform}
\Phi_\emb(x)\in\calV_\emb,\hspace{.5em}
A^\mu_\emb(x)\in\calG_\emb,\hspace{1em}x\in\bbR^{1+k},
\eeq
where $\calV_\emb$ is a vector subspace of $\calV$ and
\beq
\label{DG}
D(G_\emb)\calV_\emb=\calV_\emb. 
\eeq
Here we find the constraint (\ref{DG}) too restrictive (see the discussion
around (\ref{dd}) below). Therefore we also consider a more general set of
asymptotically embedded defects that as $x\rightarrow\infty$ have
\beq
\Phi\sim \Phi_\emb(x)\in\calV_\emb,\hspace{.5em}
A^\mu\sim A^\mu_\emb(x)\in\calG_\emb.
\eeq
These asymptotically coincide with an embedded defect, but may differ
from that form (\ref{embform}) elsewhere.

An embedded defect is a solution of the full theory if the field
equations reduce to consistent field equations 
on its embedded theory~\cite{vach94}. This requirement gives four constraints 
from the two field equations in (\ref{fe}):\\
\noindent (a) The current, evaluated from $\Phi_\emb$ and $A^\mu_\emb$,
satisfies
\beq
\label{c1}
\inprod{J^\nu\!}{\,\calG_\emb^\perp}=0
\eeq
with $\calG=\calG_\emb\oplus\calG_\emb^\perp$.
This constrains $\calG_\emb\subseteq\calG$.\\
\noindent (b) The kinetic scalar term satisfies
\beq
\label{c2}
\inprod{D_\mu D^\mu \Phi_\emb}{\calV_\emb^\perp}=0
\eeq
with $\calV=\calV_\emb\oplus\calV_\emb^\perp$. While this is the
case when (\ref{DG}) is satisfied, it is otherwise very restrictive and 
generally only holds asymptotically for certain parameter values.\\
\noindent (c) The scalar potential, evaluated for $\Phi_\emb$, satisfies
\bea
\label{scderiv}
\inprod{\pderiv{V}{\Phi}}{\calV_\emb^\perp}=0.
\label{eq-2}
\eea
This condition constrains the potential --- for instance it
holds in the BPS limit.\\
\noindent (d) The gauge kinetic terms satisfies
\beq
\label{DF}
\inprod{D_\mu F^{\mu\nu}}{\calG_\emb^\perp}=0.
\eeq
Equation (\ref{DF}) always holds
by algebraic closure of $\calG_\emb$, as observed in~\cite{vach94}.

\subsection{Embedded monopoles}

By (\ref{emb}) an embedded monopole is defined by
embedding an $\su(2)\rightarrow\u(1)$ symmetry breaking in that of
the full theory
\beq
\label{su2emb}
\begin{array}{ccc}
\calG &\rightarrow& \calH \\
\cup && \cup \\
\su(2) &\rightarrow& \u(1)
\end{array}
\eeq
Then the embedded monopole has a 't~Hooft-Polyakov form~\cite{hooft} 
on the $\su(2)$ subtheory
\bea
\label{monemb}
\Phi(\bm r) = \frac{H}{r}\,D(g(\rhat))\Phi_0,\hspace{1.2em}
A^i(\bm r) = \ds\frac{K-1}{r}\,\epsilon_{iab}\hat r_bt_a
\eea
with $\su(2)$ basis $[t_a,t_b]=\epsilon_{abc}t_c$ and 
$g(\rhat)=e^{\varphi t_3}e^{\vartheta t_2}e^{-\varphi t_3}$
in spherical polars.  Likewise, an 
asymptotically embedded monopole only has its asymptotic fields similar to 
(\ref{monemb}), where~\cite{go}
\beq
\label{asymonemb}
H-r = \O[\exp(-m_S r)],\hspace{1em}
K = \O[\exp(-m_V r)].
\eeq
Here $m_S$ and $m_V$ are the scalar and gauge boson masses in the embedded
$\su(2)$ subtheory.
Elsewhere the fields of an asymptotically embedded monopole
may differ from the embedded monopole.

Both an embedded and asymptotically embedded monopole have a long range 
magnetic field~\cite{com2}
\bea
\label{B}
-\half\epsilon_{ijk}F^{jk}
\sim\ds\frac{\hat r_i}{r^2}\, M(\rhat),\hspace{1.5em}
M(\rhat)=\Ad(g(\rhat))\,t_3.
\eea
The scalar field asymptotically tends to the vacuum with $\Phi\sim\Phi_0$ in
the $\hat x_3$-direction. Then
the magnetic generator $M=M(\hat\bm z)$ is an element of $\calH$
and satisfies the topological quantization $\exp(4\pi M)=1$~\cite{engle}.

To simplify the following calculations, we express (\ref{monemb})
in a unitary gauge (so $\calV_\emb=\bbR\,\Phi_0$).
This expression is achieved with a gauge transformation 
\bea
\Phi\mapsto D(g^{-1})\Phi,\hspace{1.5em}
\bm A\mapsto\Ad(g^{-1})\bm A - (\bm\nabla g^{-1}) g,
\eea
which takes the embedded monopole (\ref{monemb}) to~\cite{jr,comment}
\bea
\label{ungauge}
\Phi(\bm r)=\frac{H}{r}\,\Phi_0,\hspace{1.5em}
\bm A(\bm r)=-\bm A_D\,t_3 -\frac{K}{r}\,\hat\bm\eta_st_s.
\eea
Here $\bm A_D=\hat\bm\varphi(1-\cos\vartheta)/r\sin\vartheta$ is the 
Dirac gauge potential ($\bm\nabla\cdot\bm A_D=0$,
$\bm\nabla\wedge\bm A_D=\rhat/r^2$) and
\beq
\hat\bm\eta_1=\sin\varphi\,\hat\bm\vartheta
+\cos\varphi\,\hat\bm\varphi,\hspace{1em}
\hat\bm\eta_2=-\cos\varphi\,\hat\bm\vartheta
+\sin\varphi\,\hat\bm\varphi
\eeq
are two orthonormal unit vectors orthogonal to $\rhat$.
To prove (\ref{ungauge}), we write the gauge field as
\beq
\bm A = \frac{K-1}{r}\left( 
(\hat\bm\varphi\cdot\bm t)\hat\bm\vartheta
- (\hat\bm\vartheta\cdot\bm t)\hat\bm\varphi 
\right),
\eeq
then use the following identities
\bea
\Ad(g^{-1})\hat\bm\vartheta\cdot\bm t&=&
\cos\varphi\, t_1 + \sin\varphi\, t_2,\\
\Ad(g^{-1})\hat\bm\varphi\cdot\bm t&=&
-\sin\varphi\, t_1 + \cos\varphi\, t_2,
\eea
and evaluate 
\bea
\bm\nabla g^{-1}g=\bm A_D\,t_3
-\frac{1}{r}\,\Ad(g^{-1})\left( (\hat\bm\varphi\cdot\bm t)\hat\bm\vartheta
-(\hat\bm\vartheta\cdot\bm t)\hat\bm\varphi \right)
\eea

Now, our task is to find when such a monopole solution is fully or 
asymptotically embedded. This is determined by the constraints
(\ref{c1}) and (\ref{c2}).

To examine the first constraint $\inprod{J_\nu}{\calG_\emb^\perp}\!=\! 0$,
we start by rewriting (\ref{su2emb}) as
\bea
\label{decomps}
\begin{array}{ccccc}
\calG & = & \calH & \oplus\ & \calM \\
\cup && \cup && \cup \\
\su(2) & = & \u(1) & \oplus & \calN
\end{array}
\hspace{2.5em}
\calN=\bbR\,t_1\oplus\bbR\,t_2.
\eea
Substituting $J^\nu$ below (\ref{fe}) into (\ref{c1}), we find
\beq
\inprod{d(\calG_\emb^\perp)\Phi_0}{D^\nu\Phi}
=\inprod{D^\nu\Phi}{d(\calG_\emb^\perp)\Phi}=0.
\eeq
Now $D^0\Phi=0$, while spatially in a unitary gauge
\beq
D^i\Phi=(H/r)'\,\hat x^i\,\Phi_0 + (HK/r^2)\,
d(\hat\eta_s^it_s)\Phi_0.
\eeq
Using this and noting $\inprod{d(\calG)\Phi_0}{\Phi_0}=0$, we see
that (\ref{c1}) is an algebraic constraint upon
the embedding (\ref{su2emb})
\beq
\inprod{d(\calG_\emb^\perp)\Phi_0}{d(\calN)\Phi_0}
=\inprod{d(\calN)\Phi_0}{d(\calG_\emb^\perp)\Phi_0}=0.
\eeq
To this we apply a result proved in~\cite{mevor}:
\bea
\inprod{d(X_a)\Phi_0}{d(Y_b)\Phi_0} =
\la_a\la_b\inprod{X_a}{Y_b},\hspace{3em}\nn\\
\la_a=\frac{\norm{d(X_a)\Phi_0}}{\norm{X_a}},\hspace{1.5em}
X_a\in \calM_a,Y_b\in\calM_b.
\eea
Therefore if $\la_a\neq\la_b$ the monopole embedding in (\ref{su2emb}) 
is given by
\beq
\label{Nsub}
\calN \subseteq \calM_a,
\eeq
with $\calM_a$ a gauge family in (\ref{Mdecom}). If
$\la_a=\la_b$ the embedding can also be between 
gauge families (see ref.~\cite{mevor}).

For the second constraint
$\inprod{D_\mu D^\mu \Phi_\emb}{\calV_\emb^\perp}=0$, we evaluate
the scalar kinetic term in a unitary gauge. As observed above 
$D^0\Phi_\emb=D^0D^0\Phi_\emb=0$ ---
then we only need to consider the spatial components
\bea
\label{eDD}
D^iD^i\Phi_\emb&=& [(H/r)'' + 2(H/r)'/r]\,\Phi_0 \nn\\
&&\hspace{1.5em}+\,(HK^2/r^3)\, d(\hat\eta_s^it_s)d(\hat\eta_s^it_s)\Phi_0.
\eea
In evaluating this, we used the identity
\beq
\partial^i\hat\eta_s^i=\epsilon_{st}
\frac{\cos\vartheta-1}{r\sin\vartheta}\,\hat\varphi^i\eta_t^i
\eeq
and took from $d([X,Y])=[d(X),d(Y)]$,
\bea
d(t_3)d(t_1)\Phi_0=d(t_2)\Phi_0,\hspace{.5em}
d(t_3)d(t_2)\Phi_0=-d(t_1)\Phi_0.
\eea
Therefore by (\ref{eDD}), condition (\ref{c2}) is satisfied when
\beq
\label{dd}
d(t_1)d(t_1)\Phi_0 + d(t_2)d(t_2)\Phi_0 \propto \Phi_0.
\eeq
Note (\ref{dd}) is a very constraining assumption.

Therefore if both conditions (\ref{Nsub}) and (\ref{dd}) hold, there are
embedded monopole solutions like (\ref{monemb}). Towards
the core their scalar fields vanish and these monopoles are similar 
to a 't~Hooft-Polyakov monopole.

A more usual situation is when (\ref{dd}) does not hold. Before we discussed
asymptotically embedded monopoles, which coincide with an embedded monopole
only at infinity. Consequently, the question arises whether condition
(\ref{c2}) could be satisfied only in the asymptotic region. 

To answer this question, we consider the asymptotic form of (\ref{eDD}).
Substituting (\ref{asymonemb}) into (\ref{eDD}), we find
\bea
(H/r)'' + 2(H/r)'/r&=&\O[\exp(-m_S r)/r],\\
HK^2/r^3&=&\O[\exp(-2m_Vr)/r^2].
\eea
Then condition (\ref{c2}) is satisfied if the second term 
in (\ref{eDD}) is negligible
compared to the first. Exponentials beat powers, so this occurs when
the scalar mass $m_S$ is less than twice the gauge mass $m_V$. When so,
the asymptotically embedded monopoles are classified only by (\ref{Nsub}).

From now on we examine the first constraint (\ref{Nsub}). 

\subsection{Solution sets of gauge equivalent monopoles}

Because each monopole is defined from an $\su(2)$ embedding and these
embeddings generally form degenerate sets, we expect the monopoles to
have some degeneracy. Our tactic for determining this degeneracy
is to consider the action of the residual gauge symmetry $H$ upon the
$\su(2)$ embedding.

To start, consider a rigid $H$ transformation (no space-time dependence)
of a monopole's asymptotics
\beq
\label{ga1}
\Phi({\bm r}) \mapsto D(h)\Phi({\bm r}),\hspace{2em}
{\bm A}({\bm r}) \mapsto \Ad(h){\bm A}({\bm r}).
\eeq
By~(\ref{monemb}) this simply takes ${\bm t} \mapsto \Ad(h){\bm t}$, from
which the $\su(2)$ embedding moves to
\beq
\su(2) \mapsto \Ad(h)\,\su(2).
\eeq
Therefore each monopole is an element in a set of gauge-equivalent monopoles. 
This set is represented by the $H$-equivalent $\su(2)$ embeddings
\beq
\label{MQ}
\calO \cong \frac{H}{C_H(\su(2))},
\eeq
where $C_H(\su(2))\subset H$ is the centralizer of $\su(2)$ in $H$ (which
acts trivially on every element in $\su(2)$).

This set can be expressed more transparently by
noting that the denominator in (\ref{MQ}) satisfies
\beq
\label{stat1}
C_H(\su(2))\!=\! C_H(\calN),\hspace{2em}\su(2)=\u(1)\oplus\calN
\eeq
with $C_H(\calN)\subset H$ the centralizer of $\calN$ in $H$ (which acts
trivially on every element in $\calN=\bbR t_1\oplus\bbR t_2$). 
Expression (\ref{stat1}) follows from the $\su(2)$ 
commutation relation $[t_1,t_2]=t_3$: since
$[\Ad(c)t_1,\Ad(c)t_2]=\Ad(c)t_3$, then
$C_H(\calN) \subseteq C_H(\u(1))$ and (\ref{stat1}) is implied.

Next, we note that
\beq
\label{stat2}
C_H(\calN) = C_H(X),\hspace{1.5em}X \in \calN.
\eeq
This expression 
follows from again using $C_H(\calN) \subseteq C_H(\u(1))$, which implies 
$[\u(1),C_H(\calN)]=0$. Then, because any $X' \in \calN$ is 
proportional to $\Ad(h)X$ for some $h \in \U(1)$, we infer
$C_H(X')=C_H(X)$ and obtain (\ref{stat2}).

Therefore  by (\ref{stat1},\ref{stat2}) each fully or asymptotically 
embedded monopole is an element in a set
\beq
\label{mainres}
\calO \cong \frac{H}{C_H(X)},\hspace{1.5em}X\in\calN.
\eeq
For embedded 
monopoles, (\ref{mainres}) is exact and quantifies
their $H$-degeneracy. For asymptotically embedded monopoles,
(\ref{mainres}) refers only to their degeneracy at infinity and
further degeneracy may occur in the core.

Finally, we note there is part of the orbit (\ref{mainres}) 
that corresponds to spatial rotations. 
This part is found by considering the action of 
\mbox{$h(\chi) = \exp(t_3 \chi)$} on an $\su(2)$ embedding
\beq
\left( \begin{array}{c} t_1 \\ t_2 \end{array} \right)
\mapsto 
\left( \begin{array}{cc} \cos \chi & \sin \chi  \\ 
-\sin \chi & \cos \chi \end{array} \right)
\left( \begin{array}{c} t_1 \\ t_2 \end{array} \right).
\eeq
By (\ref{monemb}) this action is entirely equivalent to a spatial rotation. 
In passing, we note that this structure relates to the angular 
momentum of the monopole~\cite{jr,ang}.

\subsection{Similarities with embedded vortices}

Similar arguments have also been applied to embedded vortices~\cite{mevor}.
Such vortices are $\U(1)\rightarrow{\bf 1}$ Nielsen-Olesen vortices~\cite{nol} 
embedded according to (\ref{emb})
\bea
\label{voremb}
\begin{array}{ccc}
G &\rightarrow& H \\
\cup && \cup \\
\U(1) &\rightarrow& \bm 1
\end{array}
\hspace{3em}
\begin{array}{l}
\Phi(r,\vartheta) = f(r)D(e^{\vartheta X})\Phi_0,\\
\bm A(r,\vartheta) = \ds\frac{g(r)}{r}X\hat\bm\vartheta,
\end{array}
\eea
with $\U(1)=\exp(X\vartheta)$ and $X\in\calM$. In~\cite{mevor},
it was found that each vortex has its embedding constrained by
\beq
\label{vorcl}
X \in \calM_a,\hspace{3em}D(e^{2\pi X})\Phi_0=1
\eeq
with $\calM_a$ one of the irreducible subspaces in equation~(\ref{Mdecom}).
By similar arguments to those in sec.~4 these have degenerate
solution sets 
\beq
\label{vordeg}
\calO = \Ad(H)X \cong \frac{H}{C_H(X)}.
\eeq
For further discussion, we refer to reference~\cite{mevor}.

An interesting point is that the solution sets of 
embedded vortices and monopoles are very similar: with
obvious parallels between equations~(\ref{Nsub}) and (\ref{vorcl}) and the 
orbits (\ref{mainres}) and (\ref{vordeg}). 
Such parallels arise from the following sequence of embeddings
\beq
\label{monvor}
\begin{array}{ccccc}
\u(1) & \subseteq & \su(2)& \subseteq & \calG \\
\downarrow && \downarrow && \downarrow \\
0 & \subseteq & \u(1) & \subseteq & \calH.
\end{array}
\eeq
Also note that similar considerations apply to whether vortices
are fully embedded or asymptotically embedded --- this will be discussed 
elsewhere~\cite{todo}.

\subsection{Relation to fundamental monopoles}

In the following discussion, we restrict the scalar field to an adjoint 
representation of $G$. Then $\rank(G)=\rank(H)=r$ and one may choose
a maximal Abelian subgroup $T\subset H$ with orthonormal generators
$\{T_1,\cdots,T_r\}$.

Recall that the magnetic generator $M$ of a non-Abelian monopole
satisfies a topological constraint \mbox{$\exp(4\pi M)=1$}.
It can then be shown that these generators have a general form~\cite{gno,engle}
\beq
\label{fund-span}
M = \sum_{a=1}^r n_a \bm\be^*_{(a)} \cdot {\bm T},\hspace{2em}
\bm\be^*_{(a)}= \frac{\bm\be_{(a)}}{\be^2_{a}},
\eeq
where each $n_a$ is an integer and $\{\bm\be_{(1)},\cdots,\bm\be_{(r)}\}$ are 
simple roots. These simple roots span the set of all roots $\bm\al\in\Phi(G)$
\beq
\label{defroot}
i\, \ad({\bm T}) E_\al = {\bm \al} E_\al
\eeq
with each $\{E_\al\}$ a different root space.

A fundamental monopole is a monopole that has its magnetic generator
$M$ associated with a simple root, so that 
$M=\bm\be_{(a)}^*\cdot\bm T$ in (\ref{fund-span}). They are called fundamental
because any non-Abelian monopole can be decomposed into several such 
monopoles~\cite{wein} --- this is supported both by (\ref{fund-span}) 
and index theory methods in the BPS limit.

In reference~\cite{wein}, fundamental monopoles were seen to take an asymptotic 
Prasad-Sommerfeld form on the $\su(2)_\al$ subtheory that has generators
\bea
\label{tdef}
t^1_\al = (E_\al + E_{-\al})/\sqrt{2\al^2},\hspace{1em}
t^2_\al = -i(E_\al - E_{-\al})/\sqrt{2\al^2},\hspace{-2em}\nn\\
t^3_\al = \bm\al^*\cdot{\bm T} = \frac{\bm\al}{\al^2}\cdot {\bm T}.
\hspace{5em}
\eea
(Here each $E_{\pm\al}$ pair is normalized to 
$[E_\al, E_{-\al}] = i\,{\bm \al}\cdot {\bm T}$ so that
$[t^a_\al,t^b_\al]=\epsilon_{abc}t^c_\al$.)
Clearly, this structure is related to the $\su(2)$ embeddings in
(\ref{su2emb}), but how does the classification of these
fundamental monopoles relate to the embedding condition
$\calN\subseteq\calM_a$ in (\ref{Nsub})?
 
To answer this question, we define
\beq
\label{24}
\su(2)_\al = \u(1)_\al \oplus \calN_\al,\hspace{2em}
\u(1)_\al=\bbR\,\bm\al\cdot\bm T.
\eeq
Now, a symmetry breaking $G\rightarrow T$ (for suitable $\Phi_0$)
has a decomposition like (\ref{Mdecom}) into $\Ad(T)$-irreducible 
subspaces 
\beq
\label{Nal}
\calG = \calT \oplus \sum_{{\al} \in \Phi(\calG)} \calN_\al,
\eeq
where $\calT$ is the Lie algebra of the torus $T$.
Then the action of $\Ad(T)$ upon each $N_\al$ gives simply an $\SO(2)$
rotation $R$
\beq
\Ad(\exp({\bm\vartheta}\cdot{\bm T}))
\left(\begin{array}{c} t^1_\al \\ t^2_\al \end{array} \right)
=
R({\bm\al}\cdot{\bm\vartheta})
\left(\begin{array}{c} t^1_\al \\ t^2_\al \end{array} \right).
\eeq
Since $T\subseteq H$, then also $\Ad(T) \subseteq \Ad(H)$ and
each $\calM_a$ splits into several $\calN_\al$ components from (\ref{Nal}). 
Therefore
\bea
\label{Ndec}
\calM_a = \hspace{-.5em}
\sum_{\al\in\Phi(\calM_a)}\hspace{-1em}\calN_\al,\hspace{1.5em}
\Phi(\calM_a)=\{\bm\al:\calN_\al\subseteq\calM_a\}.
\eea

From (\ref{Ndec}), we see how the spectrum of fundamental 
monopoles relates to the constraint (\ref{Nsub}):
$\calM_a$ fragments into components $N_\al$, $\bm\al\in\Phi(\calM_a)$, of
which each component gives a (possible) fundamental monopole embedding. 
Out of each set $\Phi(\calM_a)$, an 
appropriate number of fundamental monopoles are taken~\cite{note}.

Let us now examine the monopoles that are embedded but not fundamental.
For instance, consider
\[
\begin{array}{ccc}
\,\calH \oplus \calM  &\rightarrow& \calH \\
\cup && \cup \\
\su(2)_{\al_{(1)}} \x \cdots \x \su(2)_{\al_{(p)}} &\rightarrow& 
\u(1)_{\al_{(1)}} \x \cdots \x \u(1)_{\al_{(p)}}
\end{array}
\]
with $\{\bm\al_{(1)}, \cdots, \bm\al_{(p)}\}$ mutually 
orthogonal roots. Then we interpret the diagonal $\su(2)$ embedding
\bea
t^3 = ({\bm \al}_{(b)}^* + {\bm \al}_{(c)}^*) \cdot {\bm T}, 
\hspace{1.5em}t^{1,2} = t^{1,2}_{\al_{(b)}} + t^{1,2}_{\al_{(c)}}
\eea
as relating to a combination of fundamental monopoles. 
By (\ref{Nsub}), this embedding
only gives a solution if both $\bm\al_{(b)}$ and $\bm\al_{(c)}$ 
are within the same $\Phi(\calM_a)$.

\subsection{Discussion}

To conclude, we summarize our results and make some comments:

\noindent (a) Our arguments relate to the following decomposition
\[
\calG=\calH\oplus\calM,\hspace{2em}
\calM=\calM_1\oplus\cdots\oplus\calM_n
\]
with each $\calM_a$ irreducible under $\Ad(H)$.

\noindent (b) Monopole embeddings are constructed by
\[
\begin{array}{ccccc}
\calG & = & \calH & \oplus\ & \calM \\
\cup && \cup && \cup \\
\su(2) & = & \u(1) & \oplus & \calN
\end{array}
\]
and satisfy a constraint
\[
\calN \subset \calM_a,
\]
where $\calM_a$ is any subspace in (a).

Such monopoles can either be fully or asymptotically embedded
(depending upon whether the embedded ansatz (\ref{monemb}) solves
the field equations everywhere or just asymptotically);
this is determined by the representation of the scalar field and the 
scalar and gauge boson masses.

\noindent (c) The $\su(2)$ embedding of each monopole lies in a set
\[
\calO \cong \Ad(H)\su(2) \cong \frac{H}{C_H(X)}, \hspace{2em}
X\in\calN
\]
that is formed by acting the set of rigid $H$ transformations on the 
monopole's long range magnetic field.

\noindent (d) Both the constraint in (b) and the degeneracy in (c) are
analogous to those for embedded vortices. We interpret this as a consequence
of the $\su(2)$ monopole embedding containing some $\u(1)$ 
vortex embeddings.

\noindent (e) The above relates to fundamental monopoles through
\[
\calM_a = \sum_{\al\in\Phi(\calM_a)}\hspace{-1em}N_\al,\hspace{2em}
\su(2)_\al=\u(1)_\al \oplus \calN_\al,
\]
where each $\calM_a$ fragments into components $\calN_\al$ that 
represent fundamental monopole embeddings. 

\noindent (f)
Finally, because
the tangent space to $\calO$ describes small, linear deformations that 
leave the monopole's energy unchanged,
we expect the gauge degeneracy in (c) is related
to the spectrum of zero modes. It is interesting that these 
deformations fall into multiplets of $C_H(X)$, which is itself closely 
related to the group of globally allowed gauge transformations $C_H(M)$.

{\em Part of this work was supported by King's College, Cambridge. 
I thank T.~Kibble for help with an earlier version of this paper.}


\footnotetext[1]{\ current email: n$\_$lepora@hotmail.com}

\begin{thebibliography}{99}
\bibitem{com} We have used same notation for both inner products, as the
one referred to should be clear from the context.
\bibitem{com2}
Note the physical magnetic field is scaled by the inverse couplings $f_a$
through (\ref{XY}).
\bibitem{comment}
That $\bm A\sim -\bm A_D$ is from the conventional definition of
$B^k_a=-\half\epsilon_{ijk}F^{ij}_a$; applying this gives
$\bm B_3 \sim \bm\nabla\wedge\bm A_D$.
\bibitem{note}
We expect there to be $\rank(\calM_a)$ fundamental monopoles in each
$\calM_a$, although proof is beyond the scope of this paper.
\end{thebibliography}
\end{document}